\documentstyle{mn} 

\input epsf

\def\plotone#1{\centering \leavevmode
\epsfxsize=\columnwidth \epsfbox{#1}}

\newcommand{\lesssim}{\mathrel{\hbox{\rlap{\hbox{\lower4pt\hbox{$\sim$}}}\hbox{$<$}}}}
\newcommand{\gtrsim}{\mathrel{\hbox{\rlap{\hbox{\lower4pt\hbox{$\sim$}}}\hbox{$>$}}}}
\newcommand{\beq}{\begin{equation}}
\newcommand{\eeq}{\end{equation}}
\newcommand{\beqa}{\begin{eqnarray}}
\newcommand{\eeqa}{\end{eqnarray}}

\newcommand{\taut}{\tau_{\rm T}}

\newcommand{\rc}{r_{\rm c}}
\newcommand{\Ne}{N_{\rm e}}
\newcommand{\re}{r_{\rm e}}
\newcommand{\mpr}{m_{\rm p}}
\newcommand{\ded}{\Delta E_{\rm D}}

\begin{document}

\title[Polarization of resonance lines from galaxy clusters]
{Polarization of resonance X-ray lines from clusters of galaxies}

\author[Sazonov et al.]{Sazonov S.Yu.$^{1,2}$, Churazov E.M.$^{1,2}$ and
Sunyaev R.A.$^{1,2}$\\ $^1$MPI f\"ur Astrophysik,
Karl-Schwarzschild-Str.~1, 85741 Garching bei M\"unchen, Germany\\
$^2$Space Research Institute (IKI), Profsouznaya~84/32, Moscow 117997,
Russia}
\maketitle

\begin{abstract}
As is known, resonant scattering can distort the surface-brightness
profiles of clusters of galaxies in X-ray lines. We demonstrate 
that the scattered line emission should be polarized and possibly
detectable with near-future X-ray polarimeters. Spectrally-resolved
mapping of a galaxy cluster in polarized X-rays could provide valuable
independent information on the physical conditions, in particular
element abundances and the characteristic velocity of small-scale
turbulent motions, in the intracluster gas. The expected degree of
polarization is of the order of 10\% for richest regular clusters
(e.g. Coma) and clusters whose X-ray emission is dominated by a
central cooling flow (e.g. Perseus or M87/Virgo).
\end{abstract}


\section{Introduction}

Although the intergalactic gas in clusters of galaxies is optically
thin to Thomson scattering ($\taut\lesssim 0.01$) for the
continuum photons, it can be moderately thick ($\tau\sim 1$) in the
resonance X-ray lines of highly ionized atoms of heavy elements. 
This causes the radiative transfer in such lines to be important
and gives rise to three major observable effects as previously discussed in the
literature.

First, the surface-brightness profile of a cluster in a resonance line, 
calculated in the optically-thin limit, should be distorted due to  
diffusion of photons from the dense core into the outer regions of the
cluster (Gilfanov, Sunyaev \& Churazov 1987a; Shigeyama 1998). This
must be taken into account when attempting to determine element
abundances from X-ray spectroscopic observations of galaxy clusters or
haloes of elliptical galaxies. This effect has been already considered
in application to ASCA, Beppo-SAX and XMM-Newton data
\cite{akietal97,tawetal97,moletal98,bohretal01,matetal01}.

Resonant scattering also determines the spectral profile
(which becomes saddle-shaped at small projected radii if $\tau\gtrsim
1$) of the emergent line \cite{giletal87a}. With next generation
X-ray spectrometers, combining high sensitivity with high energy
resolution ($E/\Delta E\sim 10^3$--$10^4$), it should be possible to
measure the profiles of individual lines and thus obtain a good deal of
new information on the physical conditions in the intergalactic gas. 

Finally, resonant X-ray lines from the intracluster gas should be
observable as absorption lines in the spectrum of a bright X-ray 
emitting background AGN. The strongest absorption lines can have
equivalent widths of several eV (Shapiro \& Bahcall 1980; Basko, 
Komberg \& Moskalenko 1981; David 2000). By comparing the equivalent width of
such a line and the flux from the same line seen in emission, the
distance to the cluster can be directly determined, independent of the
standard extragalactic distance scale (Krolik \& Raymond 1988; Sarazin 1989). 

However, there should be yet another effect related to the problem
under consideration. Namely, scattering in certain (as discussed
below) resonance X-ray lines causes the scattered emission to be
polarized. The recent advent of new technologies \cite{cosetal01} that
promise a drastically increased sensitivity of near-future X-ray
polarimeters has motivated us to consider this effect in detail.

\section{The model}
\label{model}

We first consider a model cluster of galaxies which contains gas
with an isothermal beta-law radial density profile \cite{cavfus76},
\beq
\Ne=N_0\left(1+\frac{r^2}{\rc^2}\right)^{-3\beta/2},
\label{dens_prof}
\eeq
where $N_0$ is the central electron number density and $\rc$ is the
core-radius. 

At plasma temperatures typical of clusters of galaxies ($T\sim
10^7$--$10^8$~K), all interesting X-ray lines have nearly Doppler 
absorption profiles whose width is determined by the velocities of
thermal and bulk (turbulent) motions, since the line natural width 
is relatively small. Indeed, the characteristic Doppler width is given by
\beqa
\ded&=& E_0\left(\frac{2kT}{A\mpr c^2}+\frac{
V_{\rm turb}^2}{c^2}\right)^{1/2}
\nonumber\\
&=& E_0\left[\frac{2kT}{A\mpr c^2}(1+1.4A{\mathcal M}^2)\right]^{1/2},
\label{dop}
\eeqa
where $E_0$ is the rest energy of a given line, $A$ is the atomic mass
of the corresponding element, $\mpr$ is the proton mass, $V_{\rm
turb}$ is the most probable turbulent velocity (we consider here
only small-scale motions for which the characteristic size of a
turbulent cell is much less than the photon mean free path and make a
simplifying assumption that the turbulent velocity distribution 
is Maxwellian), and ${\mathcal M}$ is the corresponding Mach
number. Compare, for example, the radiative width of the Ly$\alpha$
line of H-like iron ($E_0=6.97$~keV), $FWHM_{\rm
R}=\Gamma/(2\pi)\approx 0.2$~eV, with the corresponding Doppler width,
\beq
FWHM_{\rm D}=2\sqrt{\ln 2}\,\ded=6.0\left[\frac{kT}{10\,{\rm
keV}}(1+78{\mathcal M}^2)\right]^{1/2}\,\,\,{\rm eV}.
\eeq

Therefore, the total cross section for resonant scattering in a given line is 
\beq
\sigma(E)=\sigma_0
\exp\left[-\left(\frac{E-E_0}{\ded}\right)^2\right].
\label{sigma}
\eeq
Here the cross section at the line center is
\beq
\sigma_0=\frac{\sqrt{\pi}h\re c f}{\ded},
\label{sigma_0}
\eeq
where $\re$ is the classical electron radius and $f$ is the oscillator
strength of the given atomic transition.
 
Gilfanov et al. \shortcite{giletal87a} give a convenient
expression for the optical depth (from the cluster center to the
observer) at the midpoint af a Doppler-broadened resonance line:
\beqa
\lefteqn{\tau_0=\frac{\sqrt{\pi}}{2}
\frac{\Gamma(3\beta/2-1/2)}{\Gamma(3\beta/2)}N_{z,0}\rc\sigma_0}
\nonumber\\
\lefteqn{\approx 2.7\frac{\Gamma(3\beta/2-1/2)}{\Gamma(3\beta/2)}
\frac{N_0}{10^{-3}\,{\rm cm}^{-3}}\frac{\delta}{\delta_{\rm Fe, solar}}
\,i_z(T)}
\nonumber\\
\lefteqn{\times\frac{\rc}{250\,
{\rm kpc}}\frac{\sigma_0(10^7\,{\rm K}, {\mathcal M}=0)}
{10^{-16}\,{\rm cm}^{2}}\left[\frac{T}{10^7\,{\rm K}}
(1+1.4 A{\mathcal M}^2)\right]^{-1/2}
,}
\label{tau_0}
\eeqa
where $\Gamma$ represents the gamma-function, $\delta/\delta_{\rm Fe, 
solar}$ is the abundance of a given element relative to the solar
abundance of iron,  $N_{z,0}$ is the central number density of a given ion
and $i_z(T)$ is its relative abundance at temperature $T$. 

The undisturbed ($\tau_0\ll 1$) surface-brightness profile of line
emission from the beta-cluster (\ref{dens_prof}) is described by the
well-known formula
\beq
B_0(\rho)=\frac{1}{4\sqrt{\pi}}\frac{\Gamma(3\beta-1/2)}{\Gamma(3\beta)}
\rc\epsilon_0\left(1+\rho^2\right)^{-3\beta+1/2},
\label{therm_prof}
\eeq
where $\epsilon_0=N_0 N_{z,0}\Lambda(T)$ [erg cm$^{-3}$ s$^{-1}$] is
the volume emissivity of the plasma at the cluster center in a given
line, and $\rho$ is measured in units of $\rc$ throughout the text. 

\subsection{Scattering phase function and polarization}
\label{model_pol}

According to quantum mechanics, resonant scattering can be represented
as a combination of two processes \cite{hamilton47,chandra50}:
isotropic scattering (with a relative weight $w_1$) and dipole
(Rayleigh) scattering (with $w_2=1-w_1$). The first of these
components introduces no polarization, while the second does change
the polarization state of the radiation field. The weights $w_1$ and 
$w_2$ are functions of the total angular momentum $j$ of the ground
level and the $\Delta j$ ($=\pm 1$ or 0) involved in the transition.

The treatment of the radiative transfer problem of determining the
surface brightness and polarization of line emission from the beta-cluster
(\ref{dens_prof}) is relatively simple in the optically-thin 
limit, when only the contribution of the first scattering is
important. Although this approximation is generally invalid for the
cluster core region, it is always appropriate for projected radii
$\rho\gg 1$. We shall present below some analytic estimates obtained
in this approximation, additionally assuming constant
element abundances throughout the cluster as well as complete energy
redistribution in scattering. The latter means that the energy
distribution of photons after a single scattering is described by the
same Gaussian characterizing the absorption profile (\ref{sigma}).

It follows from the spherical symmetry of the model that the emergent
radiation will be polarized (provided that the given line has a dipole
scattering channel) in the direction perpendicular to a given
projected radius vector from the cluster center. We can therefore
characterize the observed emission by two energy-integrated Stokes
parameters: $Q$ and the surface brightness $B$. In the
single-scattering approximation, the latter may be represented as
$B\approx B_0+B_{\rm in}+B_{\rm out}$, where $B_0$ is the undisturbed
surface brightness (\ref{therm_prof}) and $B_{\rm in}$ and $B_{\rm
out}$ describe the contribution of emission singly scattered into and out
of our line of sight, respectively. In order to find the $Q$, $B_{\rm
in}$ or $B_{\rm out}$ for a given projection point ${\bf\rho}$ we need to
integrate the scattered emission over all positions (${\bf\rho},l$)
along the line of sight to the observer (here $l$ is the coordinate
along the line of sight).

We thus may write for isotropic scattering
\beq
B_{\rm in}(\rho)=\frac{\sigma_0}{\sqrt{2}}
\int dl\, N_z\int \frac{d\cos\theta\, d\phi}{4\pi}
I^\prime(\rho,l,\theta,\phi),\,\,\,Q(\rho)=0.
\label{bin_iso}
\eeq
For dipole scattering we have
\beq
B_{\rm in}(\rho)=\frac{\sigma_0}{\sqrt{2}}
\int dl\, N_z\int d\cos\theta\, d\phi\,\frac{3(1+\cos^2\theta)}{16\pi}
I^\prime(\rho,l,\theta,\phi)
\label{bin_ray}
\eeq
and
\beqa
Q(\rho)=\frac{\sigma_0}{\sqrt{2}}
\int dl\,N_z\int d\cos\theta\, d\phi\,\frac{3\sin^2\theta}{16\pi}
\cos(2\phi)
I^\prime(\rho,l,\theta,\phi).
\label{q_ray}
\eeqa
In the above formulae the factor $1/\sqrt{2}$ results
from the convolution of the incident (Gaussian) line energy profile
with the energy-dependent cross-section (\ref{sigma}), and $I^\prime$
is the solution of the undisturbed problem for energy-integrated
intensity (see below). We have not given above the corresponding formulae for
$B_{\rm out}$, which are similar to equations (\ref{bin_iso}) and
(\ref{bin_ray}) for $B_{\rm in}$, because we shall not use them explicitly.

\subsection{Calculations}
\label{model_calc}

Below we present some results obtained in the single-scattering
approximation using equations (\ref{bin_iso})--(\ref{q_ray}) and
append some results obtained earlier by Sunyaev \shortcite{sunyaev82} and
Gilfanov et al. \shortcite{giletal87a,giletal87b}. We then compare
these analytic expressions with the results of numerical computations
that allow for multiple resonant scatterings which become of importance when
$\tau_0\gtrsim 1$.

Our Monte-Carlo code allows one to treat an individual act of resonant
scattering in full detail, i.e. not resorting to the hypothesis of
complete energy redistribution. This is realized in a straightforward
way: for a photon propagating in a given direction ${\bf\Omega}$ with given
energy $E$ and polarization direction ${\bf m}$, first an ion is drawn
from a Maxwellian velocity distribution which has a velocity
component $({\bf v}{\bf\Omega})= c(E-E_0)/E_0$ (in the limit $v\ll
c$). Then the direction ${\bf\Omega^\prime}$ of the emergent photon is drawn in
accordance with the relevant scattering phase matrix (isotropic or
dipole), and the corresponding (Doppler-shifted) energy $E^\prime=
E_0[1+({\bf v}{\bf\Omega^\prime})/c]$ is found. 

\subsubsection{Ideal beta-cluster}
\label{model_calc_beta}

\subsubsection*{Isotropic scattering}

Gilfanov et al. \shortcite{giletal87a} have derived the surface-brightness
profile of the beta-cluster (\ref{dens_prof}) in the limit 
$\tau_0\ll 1$ for particular $\beta$ values and explicitly assuming an
isotropic phase function:
\beqa
\lefteqn{B(\beta=2/3)=\frac{\rc\epsilon_0}{8}}
\nonumber\\
\lefteqn{\times\left[
\frac{1}{(1+\rho^2)^{3/2}}-\frac{\tau_0}{\sqrt{2}}\frac{1}{(1+\rho^2)^2}
+\frac{\tau_0}{2\sqrt{2}}\frac{1}{(1+\rho^2)^{3/2}}\right],}
\label{i_iso23}
\eeqa
\beqa
\lefteqn{B(\beta=1)=\frac{3\rc\epsilon_0}{32}}
\nonumber\\
\lefteqn{\times\left[
\frac{1}{(1+\rho^2)^{5/2}}-\frac{\tau_0}{\sqrt{2}}\frac{1}{(1+\rho^2)^{7/2}}
+\frac{\sqrt{2}\tau_0}{45}\frac{13+5\rho^2}{(1+\rho^2)^3}\right].}
\label{i_iso1}
\eeqa
Note that the above expressions take into account the (negative)
contribution of emission scattered from our line of sight into other
directions.

\subsubsection*{Dipole scattering}

In this case we must substitute into equations (\ref{bin_ray}) and
(\ref{q_ray}) the solution of the undisturbed problem,
\beq
I({\bf r_0},{\bf\Omega})=\int_0^\infty ds\,\frac{\epsilon(r)}{4\pi}=
\frac{\epsilon_0}{4\pi}\int_0^\infty ds\,
\left(1+\frac{r^2}{\rc^2}\right)^{-3\beta},
\label{i0_beta}
\eeq
where the integration is done along the direction 
${\bf\Omega}=(\sin{\theta}\cos{\phi},\sin{\theta}\sin{\phi},\cos{\theta})$
and ${\bf r}={\bf r_0}+s{\bf\Omega}$. 

It proves impossible to do analytically the final integral (over the 
line of sight) in equations (\ref{bin_ray}) and
(\ref{q_ray}). However, the integration can be completed in the  
limit $\rho\gg 1$ if $\beta>1/2$. The latter condition, which means
that the total luminosity of the cluster converges when its outer radius
$r_{\rm out}\rightarrow\infty$, ensures that for a scattering site
located far (at $r\gg\rc$) from the cluster center the flux from the
core region subtending a small solid angle is much larger than the
cumulative flux of radiation coming from all other directions. Note
that $\beta>1/2$ for most observed clusters \cite{jonfor99}. The
resulting asymptotes for $\rho\gg\max(1,\tau_0^{1/(3\beta-1)})$ are

\beqa
\lefteqn{B_{\rm in}=
\frac{3}{64}\sqrt{\frac{2}{\pi}}\frac{\Gamma(3\beta/2)\Gamma(3\beta-3/2)}
{\Gamma(3\beta/2-1/2)\Gamma(3\beta/2+2)\Gamma(3\beta)}}
\nonumber\\
\lefteqn{\times
\left[\Gamma(3\beta/2+1/2)+\Gamma(3\beta/2+3/2)\right]
\rc\epsilon_0\tau_0\rho^{-3\beta-1},}
\label{bin_beta}
\eeqa
\beqa
\lefteqn{Q=\frac{3}{64}\sqrt{\frac{2}{\pi}}
\frac{\Gamma(3\beta/2)\Gamma(3\beta/2+3/2)\Gamma(3\beta-3/2)}
{\Gamma(3\beta/2-1/2)\Gamma(3\beta/2+2)\Gamma(3\beta)}}
\nonumber\\
\lefteqn{\times\rc\epsilon_0\tau_0\rho^{-3\beta-1},}
\label{q_beta}
\eeqa
\beq
\lefteqn{B_0=\frac{1}{4\sqrt{\pi}}\frac{\Gamma(3\beta-1/2)}{\Gamma(3\beta)}  
\rc\epsilon_0\rho^{-6\beta+1}.}
\label{b_beta}
\eeq
The degree of polarization can then be found as 
\beq
P=\frac{Q}{B_0+B_{\rm in}+B_{\rm out}}\approx\frac{Q}{B_0+B_{\rm in}}.
\label{p_beta}
\eeq

Note that the above formulae are applicable (for sufficiently large
$\rho$) for arbitrary values of the central optical depth $\tau_0$,
not only in the limit $\tau_0\ll 1$. Indeed, outer cluster regions are
optically thin to resonant scattering
($\tau(\rho)\sim\tau_0\rho^{-3\beta+1}$) and the amount of emission
scattered there is simply proportional (for $\beta>1/2$) to the total 
cluster luminosity dominated by emission from the central dense core,
regardless of whether this core is optically thin or thick. For this
reason it is important to include $B_{\rm in}$ in the denominator of
the expression (\ref{p_beta}), as its contribution to the observable
surface brightness can be large or even dominant. On the other hand,
the corresponding contribution from emission scattered out of our line
of sight is relatively small: $B_{\rm out}\sim\tau(\rho) B_0(\rho)\ll
B_{\rm in}$ when $\beta>1/2$.

In particular, we find from equations (\ref{bin_beta})--(\ref{p_beta}) that
\beqa
P(\beta=2/3)=\frac{(9\sqrt{2}/64)\tau_0}{1+(15\sqrt{2}/64)
\tau_0}\approx\frac{0.20\tau_0}{1+0.33\tau_0}
\nonumber\\
{\mathrm for}\,\rho\gg\max(1,\tau_0)
\label{p_b23}
\eeqa
and 
\beqa
P(\beta=1)=\frac{(\sqrt{2}/15)\tau_0\rho}{1+(\sqrt{2}/10)\tau_0\rho}\approx
\frac{0.094\tau_0\rho}{1+0.141\tau_0\rho}
\nonumber\\
{\mathrm for}\,\rho\gg\max(1,\tau_0^{1/2}).
\label{p_b1}
\eeqa

The polarization quickly vanishes when one approaches the
cluster center ($\rho\rightarrow 0$). This statement is readily proved for
$\tau_0\ll 1$, when we have 
\beq
P(\rho\ll 1,\beta=2/3)=\frac{9(3-4\ln 2)}{32\sqrt{2}}\tau_0\rho^2
\approx 0.045\tau_0\rho^2
\label{p_b23_core}
\eeq
and
\beq
P(\rho\ll 1,\beta=1)=\frac{\sqrt{2}}{15}\tau_0\rho^2
\approx 0.094\tau_0\rho^2.
\label{p_b1_core}
\eeq

We now proceed to discussing the results of our numerical
computations. It turns out that the cluster surface-brightness profile
(not the polarization!) corresponding to a given set of parameter
values is almost insensitive to the scattering phase function used,
isotropic or dipole. For this reason we shall not present below any
radial profiles corresponding to isotropic scattering and
concentrate on the dipole scattering case.

\begin{figure} 
\plotone{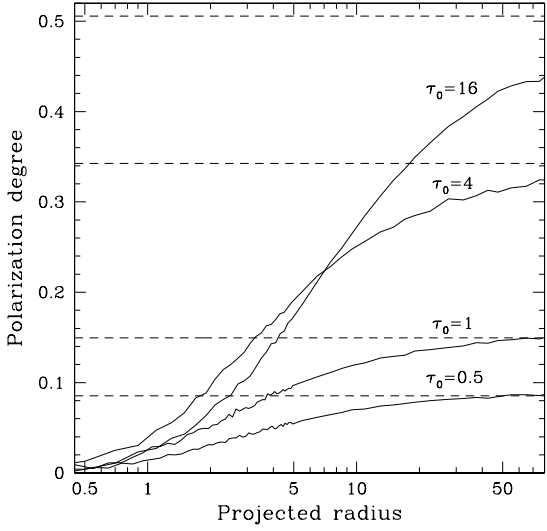}
\caption{Degree of polarization in a resonance line as a function of
projected distance from the center of a beta-cluster with $\beta=2/3$
for different optical depths at the line center (the solid
curves). The dashed curves are the corresponding
$\rho\rightarrow\infty$ asymptotes calculated from equation (\ref{p_b23}).
\label{pol_b23}
}
\end{figure}
\begin{figure} 
\plotone{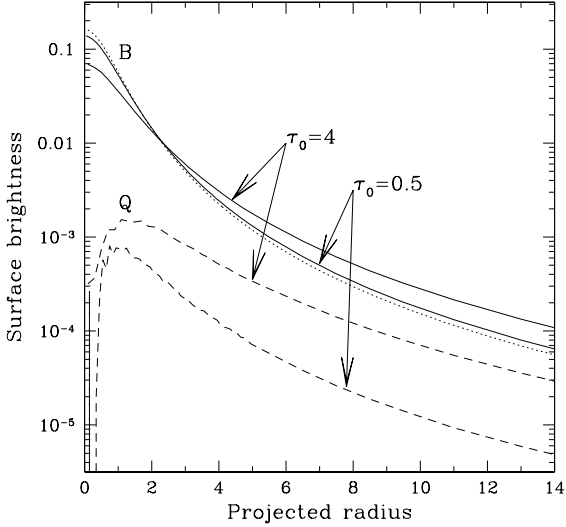}
\caption{Total surface brightness ($B$) and its polarized component
($Q$) in a resonance line as functions of projected radius  
for a beta-cluster with $\beta=2/3$ for two values of the optical
depth at  the line center. The dotted line shows the undisturbed
($\tau_0=0$) surface brightness profile. 
\label{bright_b23}
}
\end{figure}
\begin{figure} 
\plotone{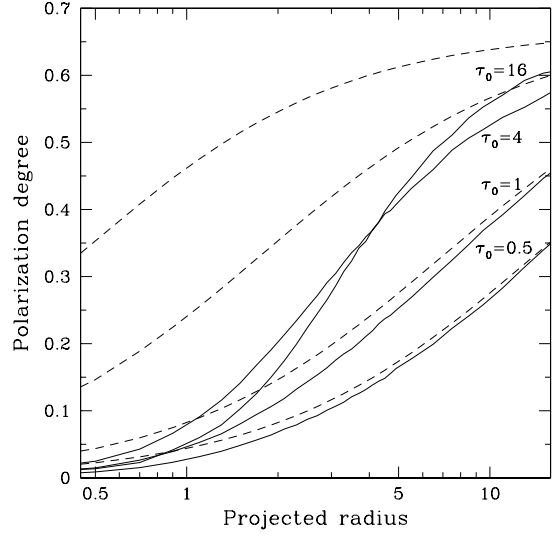}
\caption{Same as Fig.~\ref{pol_b23}, but for $\beta=1$. The dashed
curves are $\rho\rightarrow\infty$ asymptotes calculated from
equation (\ref{p_b1}).
\label{pol_b1}
}
\end{figure}
\begin{figure} 
\plotone{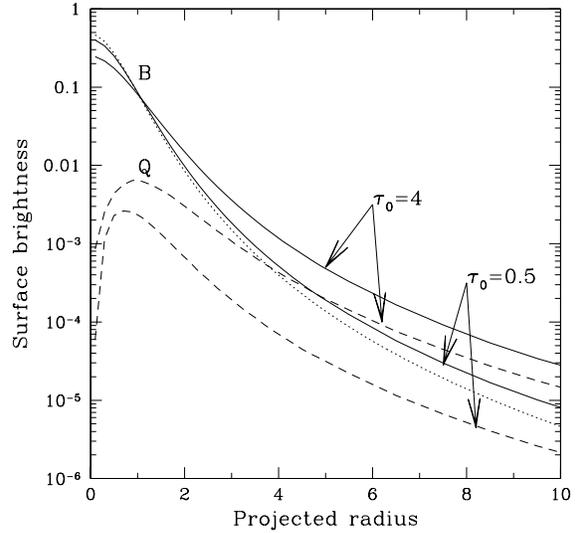}
\caption{Same as Fig.~\ref{bright_b23}, but for $\beta=1$.
\label{bright_b1}
}
\end{figure}

Figures~(\ref{pol_b23})--(\ref{bright_b1}) show a set of Monte-Carlo radial
profiles of the Stokes parameter $Q$ and surface brightness
$B$ as well as the degree of polarization $P$ for two $\beta$ values:
$2/3$ and $1$. We see that $P$ grows monotonically with increasing
$\rho$, being negligible near the cluster center and approaching at
$\rho\gg 1$ the asymptote given by equation (\ref{p_b23}) or equation
(\ref{p_b1}) if $\beta=2/3$ or $\beta=1$, respectively. In the former
case, this asymptote is a constant, while when $\beta=1$, the
polarization grows proportionally to $\rho$ until the projected distance
becomes sufficiently large that the radiation scattered into our line
of sight begins to dominate the surface brightness (see the
denominator of the expression (\ref{p_b1})). The polarization at a
given $\rho$ also grows with increasing optical depth $\tau_0$ until,
again, at a certain value of $\tau_0\sim 10$ the scattered emission
becomes prevailing and $P$ does not increase anymore. 

The effect of multiple resonant scatterings is apparent in the central
region of the cluster ($\rho\lesssim$~a few), where the degree of
polarization decreases as $\rho\rightarrow 0$ even faster than given
by the asymptotic formulae (\ref{p_b23_core}) and (\ref{p_b1_core})
when $\tau_0\gtrsim 1$.
\begin{figure} 
\plotone{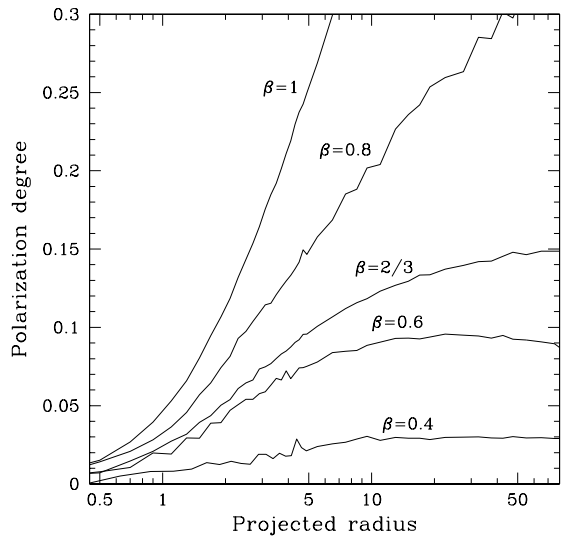}
\caption{Degree of polarization in a resonance line as a function of
projected distance from the center of a beta-cluster for different
values of the $\beta$-parameter and $\tau_0=1$. 
\label{pol_b}
}
\end{figure}
\begin{figure} 
\plotone{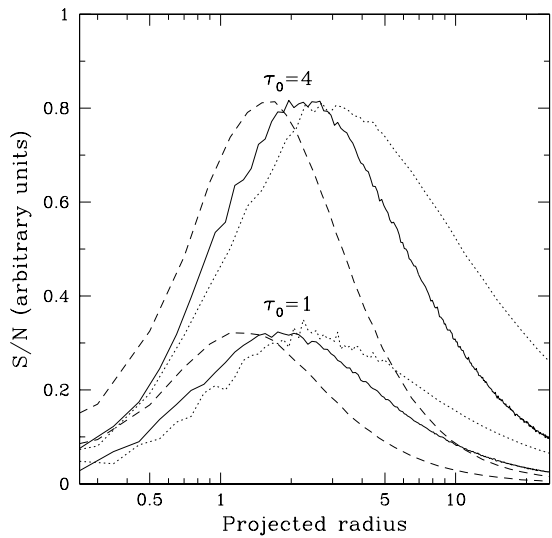}
\caption{Signal-to-noise ratio (defined as $Q/\sqrt{B}$) in detecting
polarization in a resonance line from a beta-cluster as a
function of projected radius for two valus of the central optical
depth and different $\beta$ values: $2/3$ (the solid line), 1 (the
dashed line) and 0.5 (the dotted line). 
\label{signal}
}
\end{figure}
\begin{figure} 
\plotone{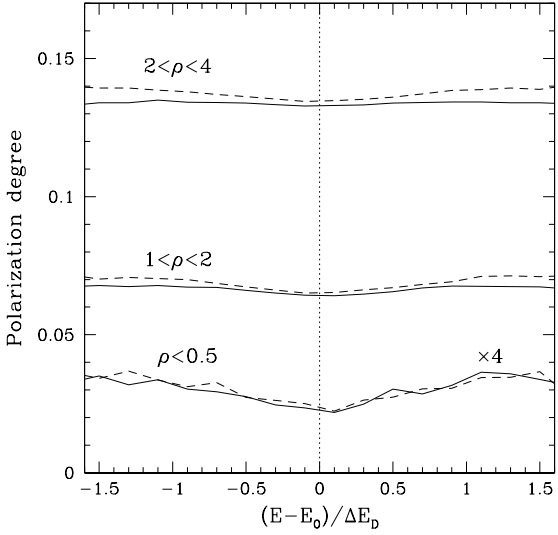}
\caption{Degree of polarization as a function of photon energy across the
spectrum of a resonance line with $\tau_0=4$ for a
beta-cluster with $\beta=2/3$ at different projected distances from the
cluster center. The solid lines result from accurate Monte-Carlo 
calculations. The dashed lines were obtained assuming complete energy
redistribution in scattering. The curves corresponding to the innermost
region $\rho<0.5$ were multiplied by 4 to make the trough
centered on $E=E_0$ visible.
\label{pol_spec_b23}
}
\end{figure}

Let us now explore how the polarization effect depends on the
$\beta$-parameter when its value is allowed to vary in a broader 
range. Fig.~\ref{pol_b} compares a set of radial profiles of $P$
obtained for different $\beta$ values and a given $\tau_0=1$. As
expected, the degree of polarization diminishes with decreasing
$\beta$ as the cluster emission becomes less centrally
peaked and the radiation field permeating the intracluster gas becomes more
isotropic. Far from the cluster core, the degree of polarization is a weak
declining function of $\rho$ when $1/2<\beta<2/3$. Indeed, in this
case $P(\rho\gg 1)\approx Q/B_0\sim\tau_0\rho^{3\beta-2}$ in accordance with
equations (\ref{q_beta})--(\ref{p_beta}). For example, $P(\rho\gg
1,\beta=0.6)\sim\tau_0\rho^{-0.2}$.

The degree of polarization remains a declining function of $\rho\gg 1$
also when $\beta<1/2$. In this case the core region no longer dominates the
radiation field within the cluster and polarization is mainly 
caused by the large-scale angular anisotropy of the radiation field. Although
this case is not easily analytically treatable, we may expect that
$Q\sim B_{\rm in}\sim \tau(\rho)B_0$ and $P(\rho\gg 1)\approx Q/B_0\sim
\tau(\rho)\sim\tau_0\rho^{-3\beta+1}$. For example, $P(\rho\gg
1,\beta=0.4)\sim\tau_0\rho^{-0.2}$, the same dependence as in the case of
$\beta=0.6$. It is clear from the profiles shown in Fig.~\ref{pol_b}
that when $\beta<2/3$, the polarization degree can be roughly described
by a constant value at $\rho\gtrsim 5$ which depends on $\beta$ and
$\tau_0$.

Given the fact that the degree of polarization is an increasing
function of $\rho$ at least within a few core radii from the cluster
center while the surface brightness falls off with moving away from
the cluster core, we may anticipate that there should exist a range of
projected radii favorable from the point of view of detectibility of the
polarized signal. In order to check this suggestion we have plotted in
Fig.~\ref{signal} the signal-to-noise ratio, which is proportional to
$Q/\sqrt{B}$, as a function of $\rho$. In doing so we, of course, assumed that 
the contribution of the instrumental and cosmic background to the
detected X-ray flux is negligible. One can see that the region $0.5\lesssim
\rho\lesssim 10$ centered (on a logarithmic scale) at $\rho_0\sim 2$
is most suitable for measurements with future X-ray polarimeters, with
the expected degree of polarization being $\sim 10\beta^2\tau_0$ per
cent at $\rho_0=2$ for $0.5<\beta< 1$ and $\tau_0< 2$.

Although in obtaining the numerical results described above we treated
the process of resonant scattering accurately, assuming complete
energy redistribution in scatterimg leads to virtually the same 
results. This is demonstrated in Fig.~\ref{pol_spec_b23}, where we
have plotted the degree of polarization as a function of photon energy
for different projected radii. Note that in the considered case of a
moderate optical depth ($\tau_0=4$), $P$ is somewhat higher in the wings of the
resonance line than at its center for the spectrum taken from the
cluster core region, while the polarization is approximately constant
across the line spectra measured at $\rho\gg 1$.

\subsubsection{Beta-cluster with a central cooling flow or AGN}
\label{model_calc_cool}

We might expect that the net degree of polarization would be
larger for clusters with a luminous cooling flow. Observed X-ray
surface brightness profiles of cooling flows usually imply an
approximately $1/r$ radial density law for the X-ray emitting gas 
which probably flattens out at $r\lesssim $ a few kpc. Therefore, when
considering a cooling-flow beta-cluster, one possibility is to
represent the cooling flow by an additional beta-model with
$\beta\approx 1/3$ and a small core-radius $\rc$. 

At the same time, it is illuminating to consider this situation as
though we have a beta-cluster with total luminosity $L_\beta$ in a given line
and a point source in the cluster center with a luminosity $L$ in the
same line. Clearly, the results of such modelling will be applicable
only at projected radii larger than the effective size of the cooling
flow (typically at $\rho\gtrsim$ a few tens of kpc). 

At the same time the results reported below are applicable to another
important situation in which there is a strong active
galaxy (AGN) in the center of a beta-cluster. In this case $L$ will
represent the luminosity of resonantly absorbed radiation from the
AGN continuum (power-law) spectrum in a given X-ray line. 

To find the contribution of the central point source to the surface
brightness and polarization in the single-scattering approximation we
must substitute 
\beq
I({\bf r_0},{\bf\Omega})=\frac{L}{4\pi r_0^2}\delta({\bf\Omega}-
{\bf r_0}/r_0),
\label{i0_point}
\eeq
into equations (\ref{bin_iso})--(\ref{q_ray}), which leads to the
results given below.

\subsubsection*{Isotropic scattering}

If $\beta=2/3$,
\beq
B_{\rm in}=\frac{\sqrt{2}}{16\pi^2}\frac{L\tau_0}{\rc^2}
\left(\frac{1}{\rho}-\frac{1}{\sqrt{1+\rho^2}}\right),\,\,\,Q=0.
\label{scat_point_iso23}
\eeq
If $\beta=1$,
\beq
B_{\rm in}=\frac{\sqrt{2}}{32\pi^2}\frac{L\tau_0}{\rc^2}
\left(\frac{\pi}{\rho}-\frac{2}{1+\rho^2}-\frac{2\arctan\rho}{\rho}\right),
\,\,\,Q=0.
\label{scat_point_iso1}
\eeq

\subsubsection*{Dipole scattering}

If $\beta=2/3$,
\beqa
\lefteqn{B_{\rm in}=\frac{3\sqrt{2}}{128\pi^2}
\frac{L\tau_0}{\rc^2}\left(\frac{3+2\rho^2}{\rho}-\frac{4+2\rho^2}
{\sqrt{1+\rho^2}}\right),}
\nonumber\\
\lefteqn{Q=\frac{3\sqrt{2}}{64\pi^2}
\frac{L\tau_0}{\rc^2}\left(\frac{1-2\rho^2}{2\rho}+\frac{\rho^2}
{\sqrt{1+\rho^2}}\right).}
\label{scat_point_dip23}
\eeqa
If $\beta=1$,
\beqa
\lefteqn{B_{\rm in}=\frac{3\sqrt{2}}{128\pi^2}
\frac{L\tau_0}{\rc^2}\left(-\frac{5+3\rho^2}{1+\rho^2}+\frac{3(1+\rho^2)}
{\rho}\arctan\frac{1}{\rho}\right),}
\nonumber\\
\lefteqn{Q=\frac{3\sqrt{2}}{128\pi^2}
\frac{L\tau_0}{\rc^2}\left(\frac{1+3\rho^2}{1+\rho^2}+\frac{1-3\rho^2}{\rho}
\arctan\frac{1}{\rho}\right).}
\label{scat_point_dip1}
\eeqa

One can see that the effect of a  particular phase function
(isotropic or dipole) on the surface-brightness profile of scattered
emission is quite small even in the case under consideration, where we
are dealing with a compact (not diffuse as in the case of a
self-illuminating beta-cluster) illuminating source. In particular if
$\beta=2/3$, $I^{\rm dip}_{\rm scat}/ I^{\rm iso}_{\rm scat}=9/8$ for 
$\rho\rightarrow 0$ and $15/16$ for $\rho\rightarrow\infty$. If
$\beta=1$, $I^{\rm dip}_{\rm scat}/I^{\rm iso}_{\rm scat}=9/8$ for
$\rho\rightarrow 0$ and $9/10$ for $\rho\rightarrow\infty$.

\begin{figure} 
\plotone{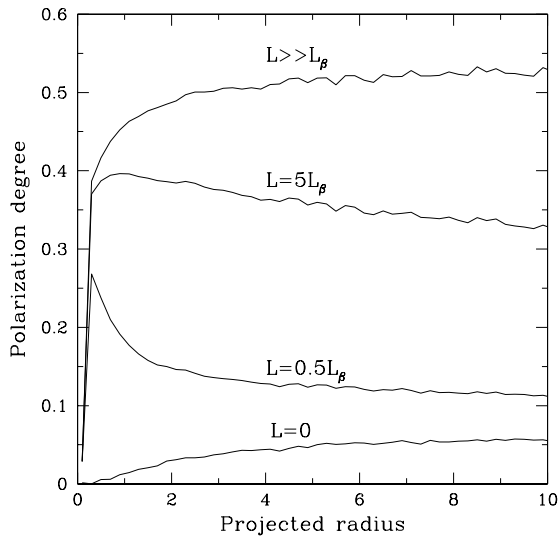}
\caption{Polarization degree in a resonance line with $\tau_0=1$ as a
function of projected radius for a $\beta=0.5$ beta-cluster with a point 
source in the center for different ratios of the luminosity of the
central source ($L$) to the total luminosity of the cluster $L_\beta$
in the line.
\label{src_pol_b0.5}
}
\end{figure}
\begin{figure} 
\plotone{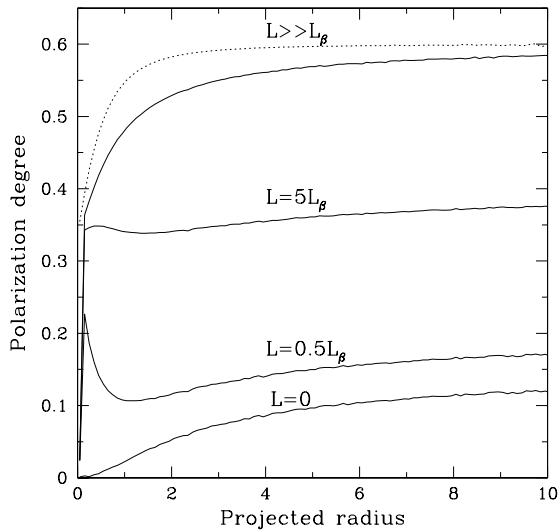}
\caption{Same as Fig.~\ref{src_pol_b0.5}, but for $\beta=2/3$. Also
shown is the analytic result (\ref{scat_point_dip23}) valid in the
small optical depth limit ($\tau_0\ll 1$).
\label{src_pol_b23}
}
\end{figure}

If we ignore for the moment the resonant scattering of the
beta-cluster emission, then the total observed surface brightness will
be $B=B_0+B_{\rm in}$ and the corresponding polarization degree
$P=Q/(B_0+B_{\rm in})$, with $B_0$ being given by equation
(\ref{therm_prof}). In the vicinity of the central source (at $\rho\ll
1$) the surface brightness will be dominated by the scattered
emission, so that $P=Q/B_{\rm in}$. In the case of dipole scattering,
$P\rightarrow 1/3$ when $\rho\rightarrow 0$ for arbitrary values of
$\beta$. The polarization at large projected radii ($\rho\gg 1$), will
depend on the ratio $B_{\rm in}/B_0\sim
(L/L_\beta)\tau_0\rho^{3\beta-2}$. This ratio is growing with distance
from the cluster center for $\beta>2/3$, and declining in the opposite
case. 

If the condition $B_{\rm in}\gg B_0$ is indeed met, then
\beq
P(\rho\gg 1)=\frac{1}{1+\Gamma(3\beta/2+1/2)/\Gamma(3\beta/2+3/2)}.
\label{p_point}
\eeq
This relation follows directly from equations (\ref{bin_beta}),
(\ref{q_beta}) and (\ref{p_beta}), which were originally derived for the ideal
beta-cluster assuming that its entire luminosity was emitted from its
center. In particular, $P(\rho\gg 1,\beta=2/3)=3/5$, $P(\rho\gg
1,\beta=1)=2/3$.

Figs.~\ref{src_pol_b0.5} and \ref{src_pol_b23} show the results of our
numerical simulations for the model of a beta-cluster with a point
central source. One can see that deep within the cluster core the degree
of polarization approaches the value $1/3$ as $\rho\rightarrow 0$, in
agreement with the analytic result, because the radiation field is completely
dominated by emission from the point source. When we move away from
the core region to $\rho\gtrsim$~a few, provided that $\beta>1/2$, we
effectively come to a situation in which there is a point central
source with total luminosity $L_{\rm tot}=L+L_\beta$ whose emission is
scattered by the beta-cluster. The resulting $P(\rho)$ profile is then
a function of $L/L_\beta$. If $\beta>2/3$, $P$ approaches as 
$\rho\rightarrow\infty$ the asymptotic value characterizing the
point-source case: $3/5$ if $\beta=2/3$ and $2/3$ when $\beta=1$. If
$\beta<1/2$, the role of the central point source becomes unimportant
at $\rho\gg 1$, similarly to the role of the beta-cluster core, as
discussed in \S\ref{model_calc_beta}.

We see from Figs.~\ref{src_pol_b0.5} and \ref{src_pol_b23} that already at
$L= 0.5L_\beta$ the observed line emission is strongly polarized across
the whole cluster, with $P$ being of the order of 15\% when $\tau_0=1$.

\section{Implications for observations}

Let us now discuss the prospects of observing the resonant scattering
polarization effect in actual clusters of galaxies. We shall begin
by discussing potentially interesting X-ray lines and then continue by
performing numerical simulations for a few clusters.

\subsection{Most promising lines}
\label{implic_lines}

For the discussion below as well as for the calculations in
\S\ref{clusters} it is important that given the very low density of the
intracluster gas, virtually all ions are in their ground atomic
state. In particular, only the lowest fine-structure sublevels are
populated.

\subsubsection{He-like $K_\alpha$ line}

The resonant transition $1s^2-1s2p (^{1}P_{1})$ at 6.7~keV of He-like
iron corresponds to one of the most prominent lines in the spectra of
rich clusters of galaxies. He-like iron is abundant in a wide range of
temperatures from 1.5 to 10~keV. Taking into account that the
transition has a large absorption oscillator strength of $\sim 0.7$,
the expected optical depth in the line can be large, and therefore this
line is one of the prime candidates for observing the resonant
scattering effect. Since the transition corresponds to the change of
the total angular momentum from 0 to 1, the phase function of the
scattering is a pure dipole. This means that this line is also
expected to have highest degree of polarization.

In the vicinity of this line there are several other lines including the
intercombination $1s^2-1s2p (^{3}P)$ and forbidden $1s^2-1s2p (^{3}S)$
lines of He-like iron and many dielectronic satellite lines. For the
forbidden line the optical depth is essentially zero, while for the
most important intercombination line ($1s^2-1s2p (^{3}P_{1})$) the
absorption oscillator strength is at least a factor of 10 lower than
for the resonant line. While many of the satellite lines have a
vanishing optical depth, some satellite lines do have an appreciable
oscillator strength. The latter correspond to the singly exited state
of particular ions of iron (e.g. the transitions $1s^22s-1s2s2p$ for 
Li-like iron). However, Li-like iron is present in sufficient amounts
only in a relatively cool (below 4 keV) plasma, which implies that in
hot clusters the optical depth in these lines will be lower than for
the resonant He line. Note also that for the satellite lines 
autoionization of electrons (from the exited state) plays a role, thus
suppressing the emission in the satellite line in the case of a
sufficient optical depth. A similar effect called ``resonant Auger
destruction'' has been considered in application to ionized
accretion disks (e.g. Ross, Fabian \& Brandt 1996). For clusters this
effect is less important because of the relatively low optical depth
and lower ratio of autoionization and radiative decay rates (at least
for relevant transitions in the Li-like iron).  

\begin{figure} 
\plotone{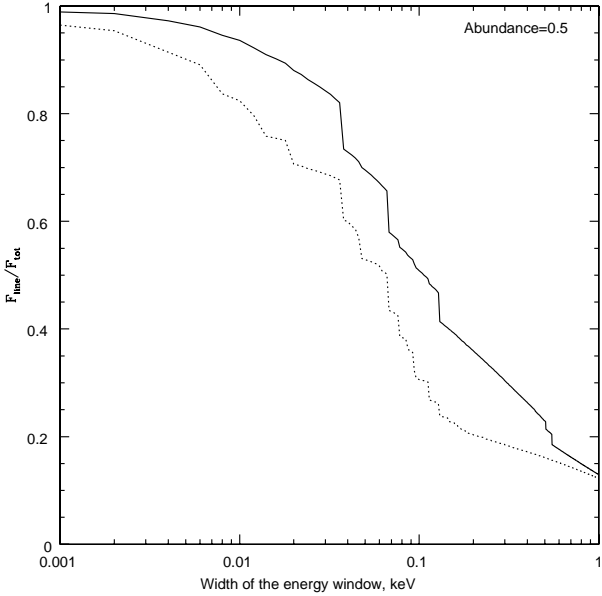}
\caption{Decrease of the observed degree of polarization for an
instrument collecting photons in a given energy range centered on the
He-like line at 6.7 keV. The solid and dotted lines correspond to a
plasma temperature of 8 and 2 keV, respectively. The abundance of iron
was assumed to be 0.5 solar. Only the flux in the line is assumed to be
polarized. The edges in the curves correspond to such widths of the energy
window when an additional prominent line falls inside the window.
\label{ldtot}
}
\end{figure}

All these lines (except for dielectronic satellites with a large 
principal number of the spectator electron in Li-like iron) are
separated from the main resonant line by at least 30~eV. Therefore,
even turbulent broadening does not severely mix the resonant line with
other lines. Finite energy resolution of the polarimeter may, however,
significantly affect the prospects of detection of the polarized
emission. Assuming that only the resonant line of He-like iron is
polarized, the measurable polarization can be estimated by multiplying the
degree of polarization as would result in the ideal case (as calculated
in \S\ref{model}) by the factor $f=F_{\rm line}/F_{\rm tot}$ where
$F_{\rm line}$ is the flux in the resonant line and $F_{\rm tot}$ is
the total flux measured by the detector in an energy band $\Delta E$
centered on the resonant line energy. In Fig.~\ref{ldtot} we plot this
factor as a function of $\Delta E$ for two values of the plasma
temperature of 8 and 2 keV. One can see that the energy resolution of a
CCD-type detector is needed to avoid a strong decrease in polarization
due to contamination by the unpolarized flux in the continuum and other lines.

\subsubsection{He-like $K_\beta$ iron line}

This line ($1s^2-1s3p (^{1}P_{1})$) at 7.88~keV is very similar to the
$K_\alpha$ line, but the oscillator strength is $\sim 0.16$, i.e. a 
factor of 4.5 smaller. This directly translates into the optical depth
being smaller by the same factor. It was suggested (e.g. Akimoto et
al. 1997; Molendi 1998) that the comparison of the radial profiles in
the $K_\alpha$ and $K_\beta$ lines may be used as an indicator of the
resonant scattering in clusters. Note, however, that the different
energies of these two lines imply some dependence of the ratio of
their emissivities on temperature. This effect may complicate an
analysis of real data.

\subsubsection{H-like $K_\alpha$ iron line}

This line at 6.96~keV has two components, $1s-2p (^{1}P_{1/2})$ and
$1s-2p (^{1}P_{3/2})$, separated by $\sim 20$~eV. The oscillator
strengths of the lines are 0.14 and 0.28 respectively. For the first
line the total angular momentum is equal to $1/2$ both for the
ground and excited states, and therefore the scattering phase function
is isotropic and scattered emission is unpolarized. For the second
line the total angular momentum of the exited state is $3/2$
and the phase function is a sum of isotropic and dipole phase
functions with equal weights. For the conditions in clusters
(i.e. only marginally optically thick cases) all complications
associated with the quantum interference of these two lines
(e.g. Stenflo 1980) can be safely neglected. We note here that the ratio of
the intensities of these two lines is essentially independent of
temperature. Therefore, a comparison of the radial profiles and
polarization properties of these lines would be the cleanest test for
the presence of resonant scattering, should next generation
calorimeters and polarimeters allow one to resolve them. However 
hydrogen-like iron is abundant only in high temperature clusters, and
the optical depth is systematically lower than for the He-like resonant line.  

\subsubsection{$K$-shell lines of lighter elements}

For low temperature clusters and especially for cooling flow regions
with the gas temperatures in the range of 1--3 keV, the K-shell lines
of lighter elements (e.g. Mg, Si or S) may be important. These
elements have abundances comparable to the iron abundance and the main
limiting factor is the ionization fraction. Compared to iron K-lines,
the cross sections (per ion) for these lines are higher -- see
equations (\ref{sigma_0}), (\ref{dop})  --  because of the lower energy of the
transition, although the smaller atomic masses imply that Doppler
broadening is more important for these elements than for 
iron (and as a consequence turbulent broadening has less impact on
these lines than on iron K-lines). 

\subsubsection{$L$-shell lines of iron}

The $L$-shell transitions (e.g. of the $2s-3p$ or $2p-3d$ type) with energies
of the order of 1 keV are especially strong at lower temperatures
of 1--3~keV typical for cooling flow regions. They are associated with
Li-, Be-, B-, C-, N-, O-, F- and Ne-like ions of iron. At any given
temperature several ionization stages are present and several strong lines are
present. Polarization properties of these transitions depend on the
total angular momenta of the ground and excited states. 


Several lines appear particularly promising and interesting for polarimetric
studies. At temperatures of the order of 1 keV this is the Be-like
iron line at 1.129 keV which has a large oscillator strength ($f=0.7$)
and a pure dipole  scattering phase function. At still lower
temperatures $\sim 0.5$ keV the most interesting is the Ne-like iron
(Fe XVII) line at 0.825 keV which has an exceptionally large
oscillator strength ($f=2.95$) and a pure dipole scattering phase
function. We note however that it is not clear if the gas in this 
temperature range is abundant in cluster cooling flows. The standard
models do predict such low-temperature plasma, while recent Chandra
and XMM observations \cite{bohretal02} do not provide clear evidence for it.

\subsection{Individual clusters}
\label{clusters}

We have performed Monte-Carlo calculations to simulate the resonant
scattering polarization effect in three prototypical clusters of
galaxies. The line energies and oscillator strengths were
taken from the list of strong resonant lines of Verner, Verner \&
Ferland \shortcite{veretal96}. The solar abundances of elements were
taken from Anders \& Grevesse \shortcite{andgre89}. We used the MEKAL code
\cite{mewetal85,mewetal86,kaastra92,lieetal95} as implemented  in the
software package XSPEC v10 \cite{arnaud96} to calculate plasma  
emissivity in a given line as a function of radius. The ionization
fractions were calculated using collisional ionization rate fits from
Voronov \shortcite{voronov97}, radiative recombination rates from
Verner \& Ferland \shortcite{verfer96} and dielectronic recombination
rates from Aldrovandi \& Pequignot \shortcite{aldpeq73}, Shull \& van Steenberg
\shortcite{shuste82} and Arnaud \& Rothenflug
\shortcite{arnroth85}. We note here that this usage of several
independent sources of atomic data may lead to certain inconsistencies
in the resulting values of the polarization degree, but we believe that the
net effect (of these inconsistencies) is very minor and can be
neglected.

\subsubsection{Coma cluster}

The Coma cluster is the prototype of a rich regular cluster. The
temperature of the gas is $kT=8.1$~keV and its density distribution
is described by the beta-law (\ref{dens_prof}) with
$\beta=0.67$, $\rc=430$~kpc and $N_0=2.3\times 10^{-3}$~cm$^{-3}$
\cite{jonfor99}. According to recent XMM-Newton 
observations the abundance of iron is 0.25 relative to the solar value
\cite{arnetal01}. At the given gas temperature He-like and H-like
ions of iron are present in nearly equal amounts with a negligible
admixture of less ionized atoms. Given the oscillator strengths of the
corresponding lines, only the 6.7~keV line of He-like iron proves to
have a substantial optical depth to resonant scattering, as indicated
in Table~\ref{coma}.

\begin{figure} 
\plotone{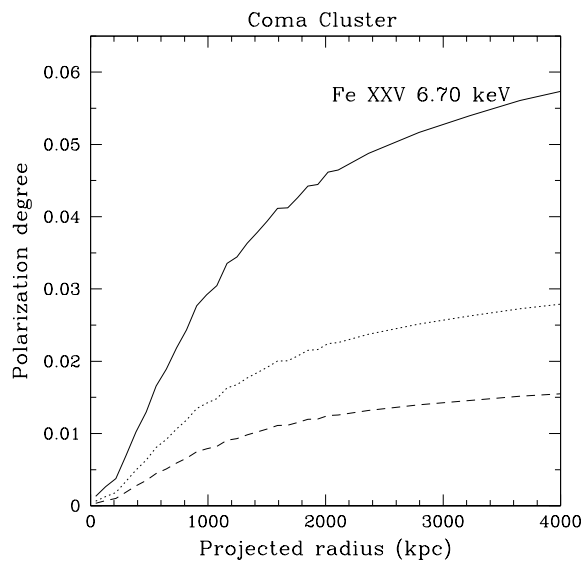}
\caption{Polarization degree in the most prominent resonant line as a
function of projected distance from the center of the Coma
cluster. The solid line corresponds to the case of negligible
turbulence in the gas, while the ratio of the characteristic turbulent velocity
to the sound velocity is ${\mathcal M}=0.2$ and ${\mathcal M}=0.4$ for
the dotted and dashed line, respectively.
\label{pol_coma}
}
\end{figure}

The reported values of $\tau_0$ were calculated assuming ${\mathcal
M}=0$. The optical depth will be smaller is we adopt a more realistic
${\mathcal M}>0$ value, because turbulent broadening is especially
important for iron, as follows from equation (\ref{dop}). For example,
$\tau_0=0.18$ for the 6.7 keV line when ${\mathcal M}=0.2$. 
Fig.~\ref{pol_coma} shows the expected radial profile of the degree of
polarization in the 6.7~keV line as would be measured by a polarimeter
with perfect spectral resolution. We see that $P\lesssim 5$\% is
expected. The polarization would be a little smaller when measured by
a detector having a finite spectral resolution, as is clear from
Fig.~\ref{ldtot}.

\begin{table}
\caption{X-ray lines with significant optical depth ($\tau_0>0.05$) to resonant
scattering. The Coma cluster. 
}
\begin{center}
\begin{tabular}{lccl}\hline \\
Ion & Energy & Optical depth & Weight of dipole \\
    & (keV)  &               & scattering       \\
\\
\hline
Fe XXIV & 1.163 & 0.04 & 0\\
Fe XXIV & 1.168 & 0.07 & 0.5\\
Fe XXV  & 6.700 & 0.37 & 1\\
Fe XXV  & 7.881 & 0.06 & 1\\
Fe XXVI & 6.952 & 0.05 & 0\\
Fe XXVI & 6.973 & 0.09 & 0.5\\
\hline
\end{tabular}
\end{center}
\label{coma}
\end{table}

\subsubsection{Perseus cluster (A426)}

The X-ray emission of the Perseus cluster is peaked on the
central dominant galaxy NGC~1275, suggesting the presence of a cooling
flow. 

We adopted simple analytic expressions for the radial dependence of
density and temperature in this cluster. The overall density
distribution, including the central cooling flow region, was
described by a single beta-model (\ref{dens_prof}) with the parameters
$N_0=0.03$~cm$^{-3}$, $\beta=0.47$ and $\rc=47$~kpc. The temperature
distribution was set to 
\beq
kT=7\frac{1+(r/r_0)^3}{2.3+(r/r_0)^3}\,\,\,{\rm keV},
\eeq
with $r_0=100$~kpc. The gas is coldest at the cluster center, where
$kT=3$~keV, and hottest in the outermost regions, reaching $kT=7$~keV.
The above distributions approximately correspond to the deprojected density
and temperature profiles derived from XMM-Newton data (to be published
elsewhere). Given the substantial azimuthal asymmetry of the 
Perseus cluster, these azimuthly averaged distributions can be used only
for crude estimates. Note that outside the cooling flow (at
$r> 300$~kpc) the density radial profile we use approximately
matches the beta-profile inferred from Einstein observations
\cite{jonfor99}, defined by values $N_0=4.1\times 10^{-3}$~cm$^{-3}$,
$\beta=0.58$ and $\rc=280$~kpc. The ratio of the two profiles is less
than 1.5 up to a distance of 4~Mpc.

We have adopted a constant abundance of iron across the cluster of 0.5
solar, as inferred from the XMM-Newton spectroscopic data.  

\begin{table}
\caption{X-ray lines with large optical depth ($\tau_0>0.5$) to resonant
scattering. The Perseus cluster. 
}
\begin{center}
\begin{tabular}{lccl}\hline \\
Ion & Energy & Optical depth & Weight of dipole \\
    & (keV)  &               & scattering       \\
\\
\hline
Fe XXIV & 1.163 & 0.8 & 0  \\
Fe XXIV & 1.168 & 1.6 & 0.5\\  
Fe XXV  & 6.700 & 3.3 & 1  \\
Fe XXV  & 7.881 & 0.5 & 1  \\
\hline
\end{tabular}
\end{center}
\label{perseus}
\end{table}

\begin{figure} 
\plotone{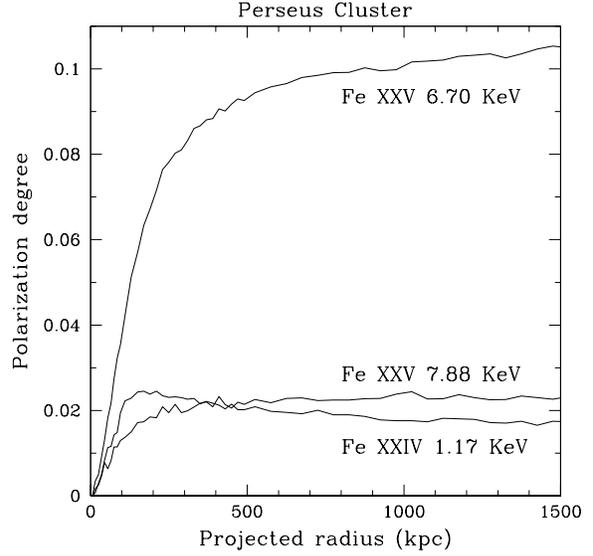}
\caption{Same as Fig.~\ref{pol_coma}, but for the Perseus cluster.
\label{pol_a426}
}
\end{figure}

Table~\ref{perseus} lists resonant lines for which $\tau_0>0.5$ (at
${\mathcal M}=0$). We see that the 6.7 keV line is again the one with
the largest optical depth. In contrast to the previous example,
Li-like iron is fairly abundant in the Perseus cluster, particularly
in its cooling flow, which gives rise to a moderate optical depth in
the Fe XXIV doublet at 1.17~keV. Fig.~\ref{pol_a426} shows the
simulated radial profiles for the degree of polarization in the lines listed in
Table~\ref{perseus}. In the case of the 1.17~keV doublet, the
polarization degree was found as $P=(Q_1+Q_2)/(B_1+B_2)$ (with
$Q_{1,2}$ and $B_{1,2}$ being the corresponding quantities for each of the
two lines). This calculation assumes that the polarimeter used cannot
resolve the lines. Note that Doppler broadening is not sufficient to
cause these lines to overlap even at sonic turbulence (${\mathcal M}=1$):
\beq
\ded=0.5\left[\frac{kT}{5\,{\rm keV}}(1+78M^2)\right]^{1/2}\,\,\,{\rm eV},
\eeq
compared with the $4.6$~eV splitting of the doublet. 

We see from Fig.~\ref{pol_a426} that the expected degree of
polarization is high for the 6.7~keV line -- larger than 8\% at
$\rho>300$~kpc. The polarization is a factor of 5 smaller for the
7.88~keV and 1.17~keV lines. This is caused by the smaller optical
depth and in the latter case also by the lower weight of dipole
scattering. We also see that the degree of polarization is nearly
constant outside the central $300$~kpc for all three sampled
lines, which is in agreement with the results of \S\ref{model}.


\subsubsection{M87/Virgo cluster}

This is another example of a cooling flow cluster, but in this
case the characteristic temperature of the cooling flow is lower ($kT\sim
1.5$~keV in the center) than in the preceeding example. For our
computations we have adopted gas distribution parameters as inferred
from XMM-Newton and ROSAT observations
\cite{nulbohr95,bohretal01,matetal01}. Namely, the density
distribution within the central 12~kpc is represented by a
beta-model with $\beta=0.47$, $N_0=0.35$~cm$^{-3}$ and $\rc=2.0$~kpc. At
larger distances, a less steep density law is used:
\beq
\Ne=0.12\left[1+\left(\frac{r}{4.3\,\mathrm{kpc}}\right)^{1.18}\right]^{-1},
\label{m87_dens}
\eeq
which is approximately a beta-law with $\beta_2=0.39$. 

The overall gas temperature distribution is approximated by
\beq
T=T_0\left[\frac{1}{1+(r/\,13.2\,\mathrm{kpc})}
+0.20\left(\frac{r}{13.2\,\mathrm{kpc}}\right)^{0.22}\right]^{-1}.
\label{m87_temp}
\eeq
The temperature rises from the center, where $T_0=1.3$~keV, towards a
broad maximum of $T=3.0$~keV at $r\sim 150$~kpc. 

The element abundances are assumed to be constant at $r<10$~kpc,
being $F(\mathrm{O})=0.4$, $F(\mathrm{Si})=1.0$, $F(\mathrm{S})=1.0$,
$F(\mathrm{Ar})=1.0$, $F(\mathrm{Ca})=1.0$ and $F(\mathrm{Fe})=0.7$
(in units of solar abundances), then gradually falling with radius to
reach the values $F(\mathrm{O})=0.4$ (unchanged), $F(\mathrm{Si})=0.6$,
$F(\mathrm{S})=0.6$, $F(\mathrm{Ar})=0.6$, $F(\mathrm{Ca})=0.6$ and
$F(\mathrm{Fe})=0.35$ at $r=40$~kpc, and remain constant from there
on. 

\begin{table}
\caption{X-ray lines with large optical depth ($\tau_0>0.5$) to resonant
scattering. The M87/Virgo cooling flow. 
}
\begin{center}
\begin{tabular}{lccl}\hline \\
Ion & Energy & Optical depth & Weight of dipole \\
    & (keV)  &               & scattering       \\
\\
\hline
O  VIII  & 0.654 & 0.4 & 0   \\
O  VIII  & 0.654 & 0.7 & 0.5 \\
Ne X     & 1.021 & 0.5 & 0   \\
Ne X     & 1.022 & 1.0 & 0.5 \\
Si XIII  & 1.865 & 2.1 & 1   \\
Si XIV   & 2.004 & 0.8 & 0   \\
Si XIV   & 2.006 & 1.7 & 0.5 \\
S  XV    & 2.461 & 2.5 & 1   \\
S  XVI   & 2.620 & 0.3 & 0   \\
S  XVI   & 2.623 & 0.6 & 0.5 \\
Ar XVII  & 3.140 & 0.8 & 1   \\
Fe XX    & 0.967 & 0.7 & 0.28\\  
Fe XX    & 0.967 & 0.5 & 0.32\\
Fe XXI   & 1.009 & 2.3 & 1   \\
Fe XXII  & 1.053 & 3.3 & 0.5 \\
Fe XXII  & 1.064 & 0.5 & 0   \\
Fe XXII  & 1.084 & 1.5 & 0.5 \\
Fe XXIII & 1.129 & 6.2 & 1   \\
Fe XXIII & 1.493 & 1.2 & 1   \\
Fe XXIV  & 1.163 & 1.6 & 0   \\
Fe XXIV  & 1.168 & 3.2 & 0.5 \\
Fe XXIV  & 1.553 & 0.6 & 0.5 \\
Fe XXV   & 6.700 & 1.8 & 1   \\
\hline
\end{tabular}
\end{center}
\label{m87}
\end{table}

Table~\ref{m87} gives a list of potentially interesting lines for
observing the resonant scattering polarization effect. There are in total
17 lines (if counting fine-structure multiplets as a single line)
with optical depth to resonant scattering exceeding 0.5. Most of these
lines, including the lines of elements lighter than Si, the lines of
He-like Si and S, and those of Fe ionic species less ionized than Fe XXIV
originate preferentially within a few tens of kpc from the M87 center
where the temperature is sufficiently low. The most interesting line
is the one at 1.129~keV, which corresponds to the $2s-3p$ transition in
Be-like iron. This line has the largest optical depth and a 
dipole scattering phase function. Also notable are the $K\alpha$
lines of helium-like Si and S at $1.865$ and $2.461$~keV,
respectively. These lines are well separated on the spectrum from the
Fe $L$-shell complex of lines and the detection of polarization in
them does not require a high spectral resolution of the polarimeter. 

\begin{figure} 
\plotone{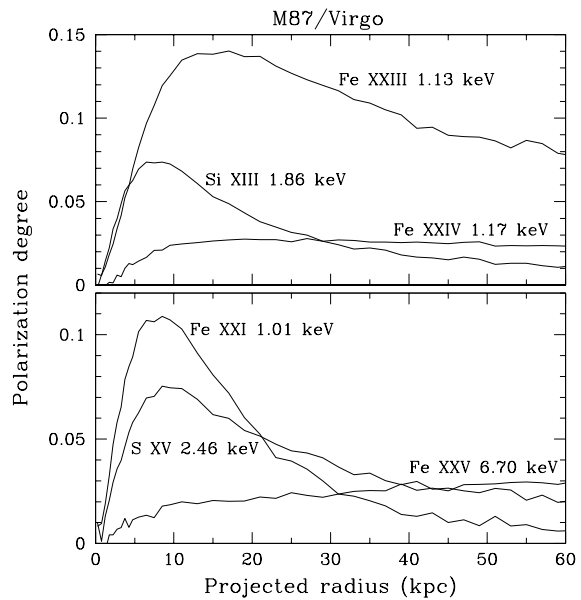}
\caption{Same as Fig.~\ref{pol_coma}, but for the M87 cooling flow.
\label{pol_m87}
}
\end{figure}

Fig.~\ref{pol_m87} shows the simulated radial
profiles of the degree of polarization for several prominent lines. As
expected, the highest polarization is achieved in the Fe XXIII 1.129~keV line,
reaching a maximum of 14\% (at ${\mathcal M}=0$) very close to the cooling flow
center (at $\rho\sim 15$~kpc), where the surface brightness is still
large. The polarization degree falls off from this maximum with moving
away from the center, and a similar behaviour pertains to the other
lines, excluding the lines of Li-like (e.g. the one at 1.17~keV) and
He-like iron (6.7~keV), which are characterized by increasing $P$
with projected radius. The case of the 1.129~keV line and other
similar lines closely corresponds to our model of a point source in
the center of a beta-cluster considered in
\S\ref{model_calc_cool}. Indeed, the fraction of the overall line
luminosity emitted within the central 10~kpc is as large as 33\% for 
the Fe XXIII 1.129~keV line, 33\% for the Si XIII 1.865~keV line, 19\%
for the S XV 2.460~keV line, 73\% (!) for the Fe XXI 1.009~keV line;
is only 15\% for the Fe XXIV 1.17~keV and as small as 2\% for the Fe XXV
6.7~keV line. 

Given the large number of resonant lines with a significant optical depth
from different elements and ionization species together
with the proximity to us, the M87/Virgo cluster appears to be a
perfect target for future X-ray polarimetric observations. It is clear
that the projected radial profiles shown in Fig.~\ref{pol_m87}
should be very sensitive to the temperature, density and abundance
physical radial profiles.

\section{Conclusions}

\begin{itemize}

\item
We have shown that the expected degree of polarization for the brightest
X-ray lines from clusters of galaxies is of the order of
10\%. 

\item
The degree of polarization is a function of projected distance 
from the cluster center, and the polarization vector is perpendicular
to the projected radius vector. This implies in particular that the
integrated X-ray flux from the whole cluster will be unpolarized
(provided the cluster if spherically symmetric). 

\item
It has also been
demonstrated that for regular clusters, whose gas density distribution
obeys the beta-law (\ref{dens_prof}), the largest signal-to-noise
ratio in measuring the polarization signal is attainable in the range
of projected radii between $0.5$ and $10$ core-radii. The degree of
polarization is zero in the direction of the cluster center, while
the photon flux from the outer parts is small making it difficult to
measure the polarization.

\item
Rich regular clusters as well as clusters with a dominant
cooling flow are the best targets for future polarimetric
observations.

\end{itemize}

Why is it important to observe the polarization effect? First of all, 
this effect is unique in the sense that there is no other obvious
mechanism that could lead to polarization in X-ray lines. Thus, a
detection of polarized diffuse emission in an X-ray line would be a
solid proof that resonant scattering is indeed in operation. 

As mentioned before, resonant scattering affects the ``apparent''
(inferred under the assumption of optically thin emission) element
abundance radial profiles of clusters
\cite{giletal87a,shigeyama98}. As a consequence, in order to 
determine the true abundance profiles from spectroscopic measurements,
one should be able to accurately subtract the distortions caused by resonant
scattering, which can well be larger in magnitude than the true
underlying radial trend. This is a difficult task, because the surface 
brightness is an integral characteristic of a given line 
of sight which passes through regions with different densities and
temperatures. One way to proceed is to compare the surface
brightness profiles of at least two spectral lines corresponding to a
given element (e.g. $K\alpha$ and $K\beta$ lines of iron, see Akimoto
et al. 1997; Molendi 1998). Polarimetric measurements could
significantly simplify such an analysis, for measuring of the
polarized flux in a single line already allows one to estimate the
contribution of resonant scattering to apparent abundance
profiles. Apart from element abundances, the resonant scattering
effects in principle can provide information on the velocity field
within the cluster, because the optical depth in a given line is a
function of the characteristic velocity of the corresponding ions. Lines of
iron are particularly sensitive to the velocity of turbulent motions.

Finally, the polarization effect can also be used for cosmological purposes. 
Indeed, since the surface brightness of resonantly scattered emission
and of its polarized component is roughly proportional to the
optical depth along a given line of sight while the surface brightness
of continuum thermal emission is proportional to the integral of the density
squared along the line of sight, it is possible to determine the
distance to the cluster and consequently the Hubble constant
analogously as is done using resonance X-ray absorption lines
\cite{kroray88,sarazin89} or the cosmic microwave background (the
Sunayev-Zel'dovich effect).

There are obvious requirements to a future X-ray polarimeter.
First, it should have the sufficient angular resolution to probe
individual parts of a cluster (we recall that the polarization
vanishes when integrated over the whole cluster). Second,
the CCD-type energy resolution is desirable in order to avoid
significant contamination of the polarized emission of a resonance line by
the unpolarized emission of neighbouring lines and the continuum. Finally,
a large effective area of the telescope is needed, larger than
is required for the detection of unpolarized resonantly scattered
emission, since the expected degree of polarization is $\sim 10$\%.  

\section{Acknowledgements}
This research was partially supported by the Russian Foundation for
Basic Research (projects 00-02-16681 and 00-15-96649) and by the
program of the  Russian Academy of Sciences "Astronomy (Nonstationary 
astronomical objects)". RS as a Gordon Moore Scholar thanks Caltech
for hospitality during the completion of this paper.


\end{document}